\documentstyle[prl,twocolumn,aps,epsf]{revtex}
\begin{document}
\twocolumn[\hsize\textwidth\columnwidth\hsize\csname@twocolumnfalse%
\endcsname
\title{A new model of quantum chaotic billiards: \\ Spectral Statistics
and Wavefunctions in 2D}

\author{E. Cuevas \cite{now}, E. Louis and J.A. Verg\'es \cite{now1}}

\address{Departamento de F{\'\i}sica Aplicada, Universidad de Alicante,
\\ Apartado 99, E-03080 Alicante, Spain.}

\date{27 November 1995}

\maketitle

\begin{abstract}
Quantum chaotic dynamics is obtained for a tight-binding
model in which the energies of the atomic levels at the boundary 
sites are chosen at random. Results for the square lattice
indicate that 
the energy spectrum shows a complex behavior with regions that
obey the Wigner-Dyson statistics and localized and quasi-ideal states
distributed according to Poisson statistics. 
Although the averaged spatial extension
of the eigenstates in the present model scales with the size of
the system as in the Gaussian Orthogonal Ensemble, 
the fluctuations are much larger.
\end{abstract}

]

\narrowtext
Recent advances in nanotechnology have made possible the fabrication
of devices in which carriers are mainly scattered by the boundaries
and not by impurities or defects located inside them \cite{1,2,3}. 
As these devices,
which are commonly referred to as quantum dots, resemble quantum
billiards, the interest in the latter has increased considerably in
the last few years. Although chaotic billiards have been intensively
investigated in the last 
twenty years \cite{4}, the behavior of their quantum analogues has not
yet been fully characterised. Some general characteristics of quantum
chaotic systems have, however, a wide acceptance. It has been shown, for 
instance, that the quantum counterparts 
of billiards having chaotic trajectories, have an energy spectrum
which obeys Wigner-Dyson statistics \cite{4,5,6}. This is the case
of the stadium and Sinai's billiards \cite{7,7a,7b,8}. On the other 
hand, it is commonly believed that there is a perfect mapping of quantum 
chaotic billiards into the more general and intensively investigated 
problem of random matrices \cite{2,7a,7b,9}.

The purpose of this Letter is to present a new model of a quantum chaotic 
billiard and investigate its spectral statistics and its 
relationship with random matrices of the Gaussian Orthogonal Ensemble (GOE). 
The model is a practical
implementation of surface roughness and its main 
characteristics 
are the following. The quantum system is described by means of a
tight-binding Hamiltonian with a single atomic level per lattice site 
in which the energies of the atomic levels at the surface
sites $S$ are chosen at random, namely,

\begin{equation}
H = \sum_{i \epsilon S}\epsilon_i c_i^{\dagger}c_i
+ \sum_{<ij>} V_{ij} c^{\dagger}_i c_j~~~~~~~, 
\end{equation}

\noindent where the operator $c_i$ destroys an electron on site $i$, and
$V_{ij}$ is the hopping integral between sites $i$ and $j$ (the symbol
$<ij>$ denotes that the sum is restricted to nearest neighbor
sites)\cite{rough}.  We take $V_{ij}=V=-1$.
The energies of the atomic levels at the boundary sites,
$\epsilon_i$, are randomly chosen between $-W/2$ and $W/2$. 
Calculations have been carried out on 
$L \times L$ clusters of the square lattice of sizes up to $L = 200$. 
Schwarz algorithm for symmetric band matrices\cite{Schwarz}
was used to compute the whole spectrum including eigenvectors for $L \le 64$
and the complete set of eigenvalues for $L \le 170$.
Instead, for larger matrices individual eigenvalues and eigenfunctions 
were obtained by inverse iteration \cite{recipes}. 

\begin{figure}
\begin{picture}(236,240) (-20,-0)
\epsfbox{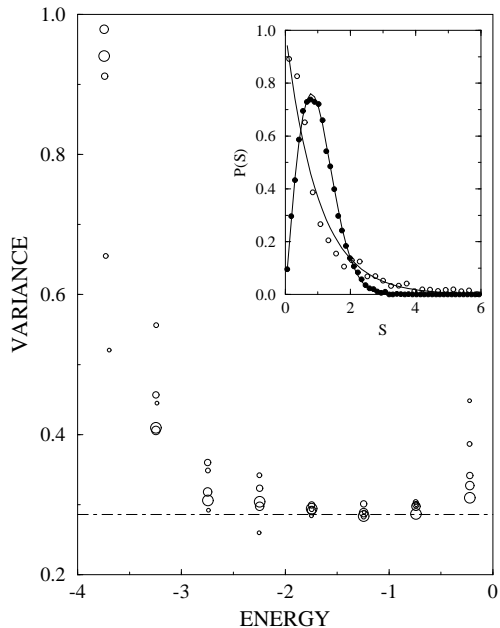}
\end{picture}
\caption{Variance of nearest-level spacings in the whole energy spectrum.
The circle size is proportional to the actual size of the system ($L$ = 
15, 30, 60, 110 and 150). The horizontal chain line indicates the 
value of the variance corresponding to the Wigner-Dyson distribution 
(0.286). Inset: Distribution of nearest level spacings in a
$170 \times 170$ cluster for energy levels between -4 and
-3.7 (empty circles) and -1.2 and -0.9 (filled circles);  
for the sake of comparison the Wigner-Dyson and the Poisson distributions 
(continuous lines) are also shown. The disorder parameter is $W=2$.
\label{fig.1}}
\end{figure}

The basis for 
expecting chaotic behavior in this model lies on the fact that the 
shift of the eigenvalues promoted by the perturbation at the surface 
would be about $(W/\sqrt{3})L^{-3/2}$ if first order perturbation theory were
applicable, which is larger than the average level separation
($\sim 8 L^{-2}$). This implies a mixture of $\sim W \sqrt{L}$ ideal
eigenstates to form a given wavefunction of the perturbed system.
A similar reasoning suggests quantum chaos for our model in any
dimension greater than 2. Note also that, in contrast with standard
chaotic billiards, our model has two length scales, namely, the size $L$
and the lattice constant $a$. Thus, even in the
the macroscopic limit ($L/a \rightarrow \infty$),
microscopic roughness remains, and is
consequently felt by quantum particles, i.e., by particles of
wavelengths of the order of $a$.

\begin{figure}
\begin{picture}(236,240) (-0,-20)
\epsfbox{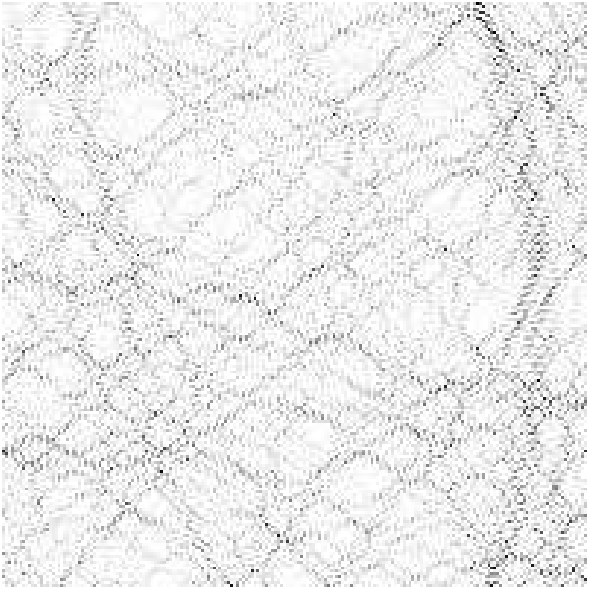}
\end{picture}
\caption{Probability amplitude of the eigenstate at $E \approx -3.3$ 
of an $200 \times 200$ cluster of the square lattice with the energies
of the atomic levels at the boundary sites randomly chosen between
-1 and 1. The probability amplitude is roughly proportional to the 
darkness.
\label{fig.2}}
\end{figure}

According to the theory of random matrices and the numerical results for
the stadium and Sinai's billiards, a clear hallmark of chaotic behavior
is a level separation statistics of the Wigner-Dyson type. Fig. 1 shows 
the variance of the nearest level spacings distribution for 
several cluster sizes of the 
billiard described by Hamiltonian (1) in the full energy spectrum. 
Away from the bottom of the band the
levels are distributed according to Wigner-Dyson statistics (this is 
explicitly shown in the inset of the figure).
The behavior is, nonetheless, somewhat different depending on
the particular energy region. In fact, whereas for energies
in the range $[-2,-0.5]$, the variance is close to that of
the Wigner-Dyson distribution (0.286) even for small clusters, 
away from that region the variance tends to 0.286 as the size is
increased. On the other
hand, near the band edges the variance of the distribution clearly 
tends to 1 (uncorrelated levels) as the system size increases, while the 
distribution approaches Poisson distribution (see inset of Fig.1). 
A similar behavior was obtained by Pavloff and Hansen in their study on
the effects of surface roughness on metallic clusters\cite{pavloff}:
in the bottom of the spectrum the de Broglie wavelength is large,
only the averaged disorder is felt and the perturbation is accordingly small.
These features differ from those of 
random matrices which show a spectrum characterised by the
Wigner-Dyson statistics throughout the whole energy range, suggesting
that, at least at a mesoscopic level, random matrices and 
chaotic billiards are not equivalent. 

\begin{figure}
\begin{picture}(236,320) (-10,-5)
\epsfbox{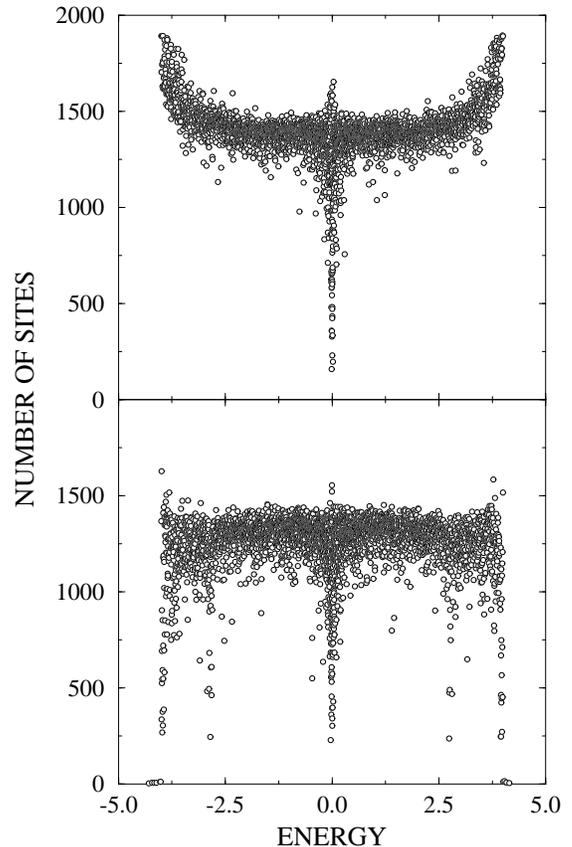}
\end{picture}
\caption{ Spatial extension of all eigenstates corresponding to
a realization of Hamiltonian (1) for a $64 \times 64$ cluster and 
two values of the disorder parameter, namely, $W=2$ (top) and $W=5$ (bottom).
\label{fig.3}}
\end{figure}

The spatial dependence of the probability density of an eigenfunction
in the energy range where chaotic behavior is expected is illustrated in 
Fig. 2. The pattern shows speckles with valleys in between, reminiscent
of the patterns found in billiards showing quantum chaotic behavior
\cite{4} and in prelocalized states in 2D \cite{10}.  
In comparing our results with those for prelocalized states 
it should be noted that in the
latter case disorder is  present at all lattice sites of the system,
whereas in our model disorder is restricted to the surface (it is in
this sense that our model can be called a billiard).
Another important
difference, which is a consequence of the previous one, is the
fact that in the present case 
Anderson localization should not be expected.

To further investigate the nature of
the states in the different energy ranges we have calculated the
participation ratio, which is a good measure of the spatial
extension of a given eigenstate. Inverse participation
numbers are defined as the moments of the distribution function
of the local weights of the eigenstates, namely,

\begin{equation}
t_{q \alpha} = \sum_{i=1}^{L \times L} |a_{\alpha i}|^{2q}~~~~~~, 
\end{equation}

\noindent where $a_{\alpha i}$ is the amplitude of the $\alpha$th
eigenstate at site $i$, i.e.,
$|\phi_\alpha> = \sum_i a_{\alpha i}|i>$.
Then the participation ratio, $P_\alpha$, is given by the inverse of 
the second moment defined by (2):

\begin{equation}
P_\alpha = t_{2 \alpha}^{-1} .
\end{equation}

\noindent $P_\alpha$ is interpreted as the number of lattice sites covered by
the eigenstate $\alpha$.
Fig. 3 shows $P_\alpha$ in the whole energy spectrum for two values of the
disorder parameter $W$. For low disorder ($W=2$)
the energy levels close to the band edges and part of the states
close to the band center are quasi-ideal.
The probability amplitude of eigenstates in that energy region 
is much like that of eigenfunctions in the fully ordered
system\cite{quasiideal}.
These states appear in energy regions in which the wavefunctions of the
ordered  cluster have small weight on the surface layer
(at band edges {\it all} states have a small surface sensibility whereas
close to the band center {\it some} of the states show small amplitudes at the
surface) whenever
the value of $W$ is small enough to allow them to keep their unperturbed
characteristics. In any case, what matters in the present
analysis is that the size of the regions of appearance of quasi-ideal states
diminishes both with the increase of $L$ and the strength $W$ of surface 
disorder.
Results of Fig 1. are consistent with this qualitative analysis.
The behavior is even more complex for large $W$ (see bottom of Fig. 3).
{\it Bona fide}
exponentially localized states appear outside the band (see circles
close to the $x$-axis in Fig.3).
Quasi-ideal states still appear at bandedges and at the band center but
in a reduced amount.
Other states show now a small spatial extent due to their character of
bulk states resonating with a particular surface impurity.
Actually, we have collected
a large sample of states showing different characteristics and allowing
therefore for different names.
To the best of our knowledge this complex
behavior has not been pointed out in previous discussions of
quantum chaotic billiards. In the limit of infinite disorder we expect
a rather simple scenario in which bulk and surface are decoupled, and,
consequently, ordered states would lie on bulk sites 
whereas localized states would be located at surface sites. 
This is a trivial limit and the most interesting
situations are of course expected for finite $W$ values. 

It is interesting to note that
a calculation for random matrices similar to that shown in Fig.3 
gives an almost energy independent distribution
with a finite width. An interesting question is how this
width (or, more precisely, this standard deviation) evolves with the 
size of the system for both random matrices and the present billiard.
In the case of
random matrices of dimension $N=L^2$ the results for the participation 
ratio averaged over
the whole energy range ($P = <P_\alpha>$) in clusters
of sizes $L=4,8,...,44,48$, can be accurately fitted by,
$P=2.2+0.33L^2$, and those for the relative standard deviation of the
distribution by $\sigma/P = 1.64/L-3.05/L^2$ (see Fig.4).
Thus, in the assymptotic limit the ratio $\sigma/P$ behaves as $1/L$. 
On the other hand, results for the present
model of quantum billiard obtained for $L=4,8,...,60,64$ and $W=2$ lead to the
following fittings: $P=0.71L+0.33L^2$ and
$\sigma/P=0.074+1.98/L-11.8/L^2+24.5/L^3$ (see Fig.4).
Averaging sets always include more than 8000 eigenstates.
These results indicate that in the macroscopic limit $\sigma/P$ 
is a constant, suggesting that fluctuations of the spatial extension of
wavefunctions are much larger than for random matrices.
The reason for this significant difference should be the surface resonances
and quasi-ideal states found in the present chaotic billiard
which seem to determine the assymptotic behavior of fluctuations
in this system. 
On the other hand we note that the fittings of the numerical results 
for $P$ show that the assymptotic behaviors of this magnitude  
for random 
matrices and for the present billiard are the same, and that significant
differences between the two models are only found for small $L$.
These results would suggest that, 
for $L \rightarrow \infty$, {\it whereas averaged properties of quantum 
chaotic billiards approach those of random matrices, 
fluctuations are much larger in the former}. 

\begin{figure}
\begin{picture}(236,170) (-10,-10)
\epsfbox{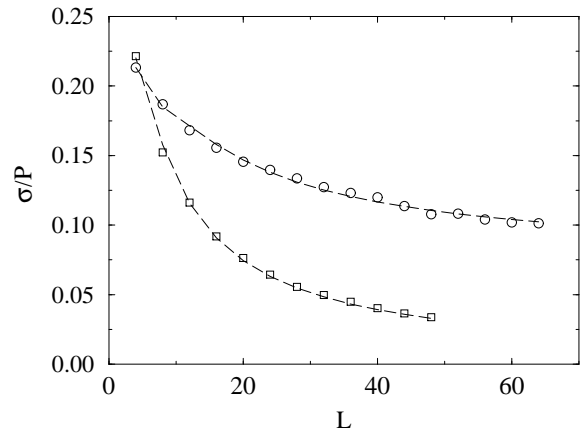}
\end{picture}
\caption{Scaling behavior of the relative fluctuation of the participation
ratio for random matrices of the GOE (squares) and for our quantum billiard
(circles). Fits are shown as dashed lines.
\label{fig.4}}
\end{figure}

The results for $P$ allow us to connect with a question of much
recent interest. We refer to the eventual multifractality \cite{11,12}
of the wavefunctions predicted \cite{13,14} and numerically calculated 
\cite{10}
for prelocalized states in disordered two dimensional systems. 
The point is whether
this exotic behavior of the eigenstates could also be a characteristic
of chaotic wavefunctions. Multifractality occurs whenever
$t_{q \alpha}\propto L^{-\tau(q)}$, $\tau(q)$ being a non-integer; 
in particular $\tau(q)=(q-1)D(q)$, where the $D(q)$ are
the generalised fractal dimensions. Our results for $P_\alpha$
(or $t_{2 \alpha}$)
averaged over the whole energy range, 
indicate that this behavior is not expected in the present case, as from
the above fittings it is concluded that $D(2)=2$. In order to ensure this
conclusion we have carried out several analysis in selected energy
ranges by means of the above method and by the standard box-counting
method \cite{12} for clusters of a fixed size. If size 
effects are properly 
accounted for, all results point to the same conclusion: the
wavefunctions of the billiard herewith investigated do not show 
multifractal behavior. In fact all results for $t_{q \alpha}L^{2q}$
can be most accurately fitted by parabollic functions.
%For instance, this is the case of $t_{\infty alpha}$, which
%can be easily calculated by taking the largest $a_{\alpha i}$ in each
%cluster.

Summarizing, we have presented a new model of quantum chaotic 
billiards which is an efficient implementation of surface roughness.
The essential feature of the model is the inclusion of diagonal
disorder at the surface of the system. The spectral statistics of 
this billiard changes through the band
in a manner not previously reported in other models of chaotic
billiards. In particular, exponentially localized, quasi-ideal,
surface resonances and chaotic states
are found to exist within the band. The probability amplitude of
chaotic eigenstates is reminiscent of that found in more standard
chaotic billiards and for prelocalized states in two dimension.  
We have also shown that whereas the assymptotic behavior of the 
the participation ratio in the present billiard is almost identical 
to that found in random 
matrices, the standard deviation (fluctuations) of that magnitude is 
much larger in the former. This
results suggests that mapping between chaotic billiards
and the random matrix problem should only be expected
for averaged properties and not for their fluctuations. 
It is very likely that these
features are not exclusive of the present billiard, and that the
behavior of quantum chaotic billiards cannot be fully described,
at least at a mesoscopic level, by random matrices.
Finally, we note that the simplicity of our model allows the study
of several situations of physical interest including the case of
3D billiards.

\acknowledgments
This work was supported in part by the spanish CICYT (grant MAT94-0058)
and DGICYT (grant PB93-1125).
Useful discussions with M. Ortu\~no and interesting
correspondence with V.I. Fal'ko are gratefully acknowledged.


\begin{references}
\bibitem{rough}
We speak about roughness because wavefunctions of well defined wavevector
are reflected by the surfaces as wavefunctions containing random amplitudes of
a large number of waves of different wavevectors whenever the wavelength of
the incoming wave is of the order of the lattice constant. 
This point of view clarifies
the classical analog of our quantum system: a square table
with imperfect sharp boundaries, i.e., surfaces that reflect incoming
particles with some random distribution of angles.
\bibitem{quasiideal}
Quasi-ideal states are those states that are marginally affected by the
existence of a disordered surface. More precisely, we call quasi-ideal states
those states that are correctly described by first-order
perturbation theory.
\end{references}
\end{document}